\documentclass[prl,letterpaper,twocolumn,preprintnumbers]{revtex4} 

\usepackage{ae}
\usepackage{bm}
\usepackage[letterpaper,colorlinks=true,pdfstartview=Fit,urlcolor=blue]{hyperref}
\newcommand{\Rref}[1]{Ref.~\cite{#1}}
\newcommand{\etal}{\textit{et~al.}}
\newcommand{\Tc}{\ensuremath{T_c}}
\newcommand{\eps}{\varepsilon}

\begin{document}

\title{Band-Structure Trend in Hole-Doped Cuprates and Correlation
  with $T_{c\,\mathrm{max}}$
}

\author{T. M. Mishonov}
\email[E-mail: ]{mishonov@phys.uni-sofia.bg}
\affiliation{Department of Theoretical Physics, Sofia University
St.~Kliment Ohridski, 5 J. Bourchier Blvd., BG-1164 Sofia, Bulgaria}
\altaffiliation{Present address: Solid State Physics Laboratory -
ESPCI, 10 rue Vauquelin, F-75231 Paris Cedex 05, France}

\noindent
\textbf{Comment on ``Band-Structure Trend in Hole-Doped Cuprates and
 Correlation with $\boldsymbol{T_{c\,\mathrm{max}}}$''}\\

In the Letter by Pavarini~\etal~\cite{Pavarini:01} a strong
correlation is observed between $T_{c\,\mathrm{max}}$ and a single
parameter $s(\eps)=(\eps_s-\eps)(\eps-\eps_p)/(2t_{sp})^2$, which is
controlled by the energy of the Cu $4s$ orbital $\eps_s$, $\eps_p$ is
the O $2p$ single cite energy, $t_{sp}$ is the transfer integral
between neighbor Cu $4s$ and O $2p$ orbitals, and $\eps(\mathbf{k})$ is
the conduction band energy whose bottom at the $\Gamma$ point
coincides with the Cu $3d_{x^2-y^2}$ energy $\eps_d$. It is unfortunate
that theorists have not so far paid any attention to this observation
because it is an important correlation between the \textit{ab initio}
calculated parameter $r=(1+s)/2$ and the experimentally measured
$T_{c\,\mathrm{max}}$ which can reveal the subtle link between the
experiment and the theory and finally solve the long-standing puzzle
of the mechanism of high-\Tc\ superconductivity (HTSC).  The purpose
of the present Comment is to emphasize that the missing link has
already been found, and the work by Pavarini~\etal\ can be used as a
crucial test for theoretical models of HTSC.  Perhaps the simplest
possible interpretation, though one could search for alternatives, is
given within the framework of the Cu $4s$-Cu $3d$ two-electron exchange
theory of HTSC~\cite{Mishonov:03}. In order for the
Schubin-Zener-Kondo exchange amplitude $J_{sd}$ to operate as a
pairing interaction of the charge carriers, it is necessary the Cu $4s$
orbital to be significantly hybridized with the conduction band. The
degree of this hybridization depends strongly on the proximity of the
Cu $4s$ level to the Fermi level $\eps_F$. Thus, it is not surprising
that $\eps_s$ controls the maximal critical temperature
$T_{c\,\mathrm{max}}$, being the only parameter of the CuO$_2$ plane
which is essentially changed for different cuprate superconductors.

Cu $3d$ and Cu $4s$ are orthogonal orbitals and their hybridization is
indirect. First, the Cu $3d_{x^2-y^2}$ orbital hybridizes with the O
$2p_x$ and O $2p_y$ orbitals, then the O $2p$ orbitals hybridize with
Cu $4s$. As a result we have a ``$3d$-to-$4s$-by-$2p$'' hybridization
of the conduction band of HTSC cuprates which makes it possible the
strong \emph{antiferromagnetic} amplitude $J_{sd}$ to create pairing
in a relatively narrow Cu $3d_{x^2-y^2}$ conduction band. The
hybridization ``filling'' of the Cu $4s$ orbitals can be seen in
cluster calculations as well~\cite{Stoll:03}. The $s(\eps)$ parameter
introduced in \Rref{Pavarini:01} reflects the proximity of all 3
levels in the generic 4-band Hamiltonian of the CuO$_2$ plane---their
``random coincidence'' for the Cu-O combination. Suppose that those
levels are not so close to each other. In this case the slightly
modified parameter
\begin{equation}
s'(\eps)=(\eps_s-\eps)(\eps-\eps_p)/(4t_{sp}t_{pd})
\end{equation}
is simply the energy denominator of the perturbation theory which describes
the hybridization filling of the axial orbital, see Eq.~(4.10) in
\Rref{Mishonov:03}. Whence $[s'(\eps)]^2$ is a denominator of the pairing
amplitude in the BCS equation Eq.~(6.4) in \Rref{Mishonov:03}. Hence, we
conclude that the correlations reported in \Rref{Pavarini:01} are simply the
correlations between the critical temperature \Tc\ and the dimensionless BCS
coupling constant $\rho(\eps_F)J_{sd}/[s'(\eps_F)]^2$. Of course, for coupling
constants $\sim 1$ the BCS trial wavefunction can be used only for qualitative
estimates, but knowing the Hamiltonian the mathematical problem may somehow be
solved. In any case, even qualitatively, we are sure that the stronger pairing
amplitude $J_{sd}$ and hybridization $1/s'(\eps_F)$ enhance \Tc.

Having LDA calculations for the band structures of many cuprates it is
worthwhile to perform a LCAO fit to them~\cite{Andersen:95} and using
experimental values of \Tc\ to extract the pairing amplitude $J_{sd}$
for all those compounds. The \textit{ab initio} calculation of the
Kondo scattering amplitude parameterized by $J_{sd}$ is an important
problem which has to be set in the agenda of computational solid state
physics. We expect that it will be a weakly material dependent
parameter of the order of the $s$-$d$ exchange amplitude in Kondo
alloys, but perhaps slightly bigger as for the Cu ion the $3d$ and
$4s$ levels are closer compared to many other ions. Closer energy
levels, from classical point of view, imply closer classical periods
of orbital motion which leads to some ``resonance'' enhancement of the
exchange amplitude due to intra-atomic two-electron correlations. The
final qualitative conclusion that can be extracted from the
correlations reported by Pavarini~\etal\ is the explanation why only
the CuO$_2$ plane renders HTSC possible, whereas hundreds other
similar compounds are not even superconducting, or have only a
``conventional'' value of \Tc. The natural explanation is: because its
$s$-parameter is not small enough below its critical value. Even among
the cuprates one can find compounds with ``conventional'' values of
\Tc\ having relatively large value of the $s$-parameter. For other
transition metal compounds the parameter $s'(\eps_d) =
(\eps_s-\eps_d)(\eps_d -\eps_p)/(4t_{sp} t_{pd})$ is much bigger than
its critical value $s_c$ which can be reached probably only for the
Cu-O combination. Thereby, the correlation reported by Pavarini~\etal\
is a crucial hint which of the models for HTSC is still on the arena.\
\\

\noindent
T. M. Mishonov

Department of Theoretical Physics, \href{http://www.phys.uni-sofia.bg}{Faculty of Physics}

\href{http://www.uni-sofia.bg}{Sofia University St.~Kliment Ohridski}

5 J. Bourchier Boulevard, BG-1164 Sofia, Bulgaria

\end{document}